\documentclass[12pt]{iopart}
\usepackage{graphicx}
\usepackage{amssymb}
\usepackage[english]{babel}
\usepackage{psfrag}
\newcommand{\bq}{\begin{equation}}
\newcommand{\eq}{\end{equation}}
\newcommand{\ba}{\begin{eqnarray}}
\newcommand{\ea}{\end{eqnarray}}
\newcommand{\dt}{\Delta t}
\newcommand{\at}{\tilde{a}}
\newcommand{\bt}{\tilde{b}}

\usepackage[T1]{fontenc}
\usepackage[cp1250]{inputenc}
\usepackage{graphicx}
\usepackage{psfrag}
\usepackage{color} 
\usepackage{bbm}
\usepackage{hyperref}

\usepackage{ifthen}

\newcommand{\pd}[2]{
	\ifthenelse{\equal{#2}{1}}{\frac{\partial}{\partial #1}}
	{\frac{\partial ^ #2}{\partial #1 ^#2}}
}
\newcommand{\dd}[2]{
	\ifthenelse{\equal{#2}{1}}{\frac{{\rm d}}{{\rm d} #1}}
	{\frac{{\rm d}^ #2}{{\rm d} #1 ^#2}}
}

\newcommand{\pf}[3]{
	\ifthenelse{\equal{#3}{1}}{\frac{\partial #1 }{\partial #2}}
	{\frac{\partial ^ #3 #1 }{\partial #2 ^#3}}
}
\newcommand{\df}[3]{
	\ifthenelse{\equal{#3}{1}}{\frac{{\rm d} #1}{{\rm d} #2}}
	{\frac{{\rm d} ^ #3 #1}{{\rm d} #2 ^#3}}
}

\begin{document} 
\title{Condensation in models with factorized and pair-factorized stationary states}
\author{M. R. Evans and B. Waclaw}
\address{
SUPA, School of Physics and Astronomy, University of Edinburgh, Peter Guthrie Tait Road, Edinburgh EH9 3FD, United Kingdom}
\eads{\mailto{martin@ph.ed.ac.uk}, \mailto{bwaclaw@staffmail.ed.ac.uk}}

\begin{abstract}
Non-equilibrium real-space condensation is a phenomenon in which a finite fraction of some conserved quantity (mass, particles, etc.) becomes spatially localised. We review two popular stochastic models of hopping particles that lead to condensation and whose stationary states assume a factorized form: the zero-range process and the misanthrope process, and their various modifications. We also introduce a new model - a misanthrope process with parallel dynamics - that exhibits condensation and has a pair-factorized stationary state.
\end{abstract}
Date \today
\maketitle

\section{Introduction}\label{intro}

Real space condensation is a non equilibrium phase transition which
occurs in various contexts such as granular clustering, traffic jams, wealth
condensation or simulations of polydisperse hard spheres \cite{burda_wealth_2002,chowdhury_statistical_2000,oloan_jamming_1998,evans_condensation_2010}.  In all of
these systems there is some conserved quantity (mass, wealth or volume
for example) which is transported through the system.  If the global density of this quantity is above a critical value, a finite fraction  condenses onto a single
lattice site or localized region of space.

Surprisingly, many features of condensation are captured within a simple
lattice model known as the zero-range process (ZRP) (for review see \cite{evans_nonequilibrium_2005, godreche_urn_2007} ).
In the simplest  one-dimensional asymmetric version of this model, a particle moves from site $i$ to $i+1$ of a one-dimensional periodic lattice with rate
$u(m)$, where $m$ is the occupancy (number of particles) of the departure site $i$.
As the rates are totally asymmetric a current always flows  and detailed balance cannot be satisfied,
thus the stationary state is non-equilibrium.
The great advantage of this model is that its non-equilibrium stationary state  has a simple
factorized form which is amenable to exact analysis.
The structure for  $P(\{m_i\})$, the probability that each site $1 \leq i\leq L$ contains mass $m_i$ is
\bq
P(\{m_i\})= \frac{1}{Z_L}\prod_{i=1}^L f(m_i)\, \delta_{M, \sum_j m_j}\;. \label{fullfact}
\eq
Thus the numerator in (\ref{fullfact}) contains one (non-negative) factor $f(m_i)$ for each site
$i$ and $f(m)$ is known as the  single-site  weight function and depends on $u(m)$ as
\bq
f(m) = \prod_{i=1}^m \frac{1}{u(i)} \quad \mbox{for}\quad m\geq 1\quad\mbox{and}\quad f(0) =1. \label{zrp_f_u}
\eq
The denominator $Z_L$ is the normalization or nonequilibrium partition function
\bq
{Z_L}= \sum_{\{m_i=0\}}^{\infty} \prod_{i=1}^L f(m_i)\, \delta_{M, \sum_j m_j}. \label{norm}
\eq
In (\ref{fullfact}) and (\ref{norm}) the Kronecker delta imposes the constraint that the total mass (or number of particles) $M$ in the system is conserved.

It turns out that the zero-range process is not the only model leading to the factorized steady state (\ref{fullfact}). In recent years, it has been shown that other models in which the hop rate depends not only on the occupation of the departure site but also on other variables can also exhibit steady states that factorize exactly
\cite{gabel_facilitated_2010,vafayi_weakly_2014, chleboun_condensation_2013,chleboun_dynamical_2015,godreche_condensation_2012,daga_phase_2015,cao_dynamics_2014} or approximately \cite{hirschberg_motion_2012} over the sites of the system. It has also been discovered that there exist various  classes of models in which the steady state factorizes over pairs of sites \cite{evans_interaction-driven_2006,waclaw_tuning_2009,ehrenpreis_numerical_2014},
\bq
P(\{m_i\})= \frac{1}{Z_L}\prod_{i=1}^L g(m_i,m_{i+1}) \delta_{M, \sum_j m_j}. 
\label{pairfact}
\eq 
In (\ref{pairfact}) $g(m_i,m_{i+1})$ is the pairwise weight.
In the following we will first briefly review some of the models with fully factorized and pair-factorized steady states and discuss conditions under which condensation occurs in these models. We shall then introduce a new model, a discrete time variant of the misanthrope process \cite{cocozza-thivent_processus_1985}, with a pair-factorized steady state in which the hopping rate depends on the occupations of the departure and arrival sites.
Such discrete time schemes are
often used in the simulation of traffic flow and pedestrian dynamics.
In  the zero-range process it has been shown that a generalisation 
to discrete time dynamics still results in a factorized stationary state
\cite{evans_factorized_2004}. In this work we show that the misanthrope process may have a
pair-factorized stationary state under discrete time dynamics and we
establish conditions under which this holds.  Further, we present an
analysis of condensation in such pair factorized states.

\section{Zero-range process}
The zero-range process is specified by the hop rate $u(m)$ which determines the properties of the steady state through the single-site weight $f(m)$ from Eq.~(\ref{zrp_f_u}). It is important to note  that any exponential factor $A q^m$ in $f(m)$ does not change the steady-state properties since it appears in Eq.~(\ref{fullfact}) as a constant prefactor $A^Lq^{\sum_i m_i}=A^L q^M$ due to the fixed total mass $M$
and number of sites $L$.  From now on we will generally suppress such exponential factors in $f(m)$.

Condensation  in models with factorized stationary states occurs in the limit $L\to\infty$ and fixed density of particles $\rho=M/L$ when the asymptotic (large $m$) behaviour of
$f(m)$ (modulo any exponential factors) is the following:
\begin{description}
\item[I] $f(m)\sim m^{-\gamma}$  with $\gamma>2$.  The critical mass density
above which condensation occurs is 
finite but its numerical value depends on the particular form of $f(m)$ and 
not only on its asymptotic behaviour. The fraction $\rho/\rho_c-1$ of all 
particles goes into the condensate. We will refer to this behaviour as standard condensation. \footnote{Actually condensation also occurs if $f(m)$ decays more quickly than $1/m^\gamma$, e.g., as a stretched exponential.}

\item[II] $f(m)$ increases with $m$ more quickly than exponentially, e.g., 
as $\sim m!$. This leads to so called strong (or complete) condensation - 
the critical density $\rho_c=0$ and a fraction of particles
tending to one in the thermodynamic limit
is located at one  site.
\end{description}
It can be shown that standard condensation ({\bf I}) occurs in the ZRP
when the hop rates in the limit of large $m$ asymptotically approach some positive value $\beta$ as
\bq 
	\frac{u(m)}{\beta} \sim 1+ \frac{\gamma}{m}
\eq
with $\gamma>2$ , or more slowly than $1/m$.
On the other hand strong  condensation occurs when  $u(m)\to 0$ as $m\to \infty$. For example, $u(m) =1/m$ yields $f(m)= m!$.

To see why the condensation happens in the two  generic cases highlighted above, we shall follow a standard approach \cite{evans_nonequilibrium_2005}. Treating the steady-state probability as the statistical weight of a given configuration, and defining the grand-canonical partition function
\bq
	G(z) = \sum_{\{m_i\}} z^{\sum_i m_i} P(m_1,\dots,m_L) = \sum_{\{m_i\}} \prod_{i=1}^L f(m_i) z^{m_i} = F(z)^L,
\eq
where
\bq
	F(z) = \sum^{\infty}_{m=0} f(m) z^m, \label{Fz}
\eq
we see that the phase transition, signaled by a singularity  of $G(z)$ at some $z_c$, is possible only if the series $F(z)$ has a finite radius of convergence $z_c$. Moreover, the density calculated as a function of fugacity $z$ from the grand-canonical partition function:
\bq
	\rho(z) = z \frac{F'(z)}{F(z)}
\eq
must yield  a  value $\rho_c < \infty$  as $z  \nearrow z_c$ so that
the singularity in $G(z)$ and  accompanying phase transition occur at finite density.
 This is only possible if either $f(m)$ 
decays as a power law in which case we may have a finite $\rho_c$ (case {\bf I})
or $f(m)$ grows very fast with $m$ in which case $z_c=0$ and $\rho_c=0$ (case {\bf II}). 

Thus the grand canonical ensemble can only realise  densities $\rho \leq \rho_c$. When $\rho >\rho_c$ one must work in the canonical ensemble
(fixed number of particles) \cite{evans_nonequilibrium_2005,evans_canonical_2006}. It turns out that the excess  mass $M_{\rm ex} = M-L\rho_c$ condenses onto a randomly selected lattice site and forms the condensate. The remainder of the system (referred to as the fluid) is described by the grand canonical ensemble at the critical density $\rho_c$ \cite{groskinsky_condensation_2003,majumdar_nature_2005,evans_canonical_2006,armendariz_conditional_2011,armendariz_zero-range_2013}.

\section{Generalized Class of Models with Factorized Stationary State\label{Sec:ZRPgen}}
So far we have discussed the ZRP as  an example of a simple model with a factorized stationary state
(\ref{fullfact}). More generally one can ask,  when does a stochastic mass transport model have a such
a factorized stationary state if the hop rate depends only on the state of the departure site? To this end a class of models  was studied in \cite{evans_factorized_2004}
that generalises the ZRP in a number of different ways while maintaining the factorization of the steady state.
First, more than one unit of mass can be transferred from site $i$ to $i+1$.
Second  the dynamics consists of discrete time  update and
simultaneous transport of mass at different locations is possible at an update.
More precisely,  in each time step, some number
$0 \leq \mu_i\leq m_i$ of particles depart from site $i$ and move to site $i+1$ with
probability $\phi(\mu_i|m_i)$ which is known as the chipping kernel.  For conservation of probability we
require $\sum_{\mu=0}^m \phi(\mu|m,n)=1$.  

It was shown \cite{evans_factorized_2004} that
a necessary and sufficient condition for the stationary state of this class of models to factorize is that the chipping kernel takes the form
\bq
	\phi(\mu|m) = \frac{u(\mu) v(m-\mu)}{f(m)}, \label{zrp_nufact}
\eq
where the single site weight $f(m)$ is given by
\begin{equation}
f(m)=\sum_{\mu=0}^m u(\mu)v(m-\mu)\;.
\end{equation}
In these expressions $u$ and $v$ are  positive functions.
Expression (\ref{zrp_nufact}) implies a factorization of the chipping kernel
into a factor $u(\mu)$ which depends on the mass transferred
and a factor $v(m-\mu)$ which depends on the mass which remains.
The model can be further generalized to a continuous mass variable \cite{evans_factorized_2004} but we shall not consider that here.
The  model was also considered on an arbitrary graph rather than a periodiic chain and a condition similar to (\ref{zrp_nufact}) was show to be sufficient for factorization \cite{evans_factorized_2006}.

\section{Misanthrope process}
As already stressed the defining feature of the ZRP and the class of models just discussed
is that transition rates or probabilities for the transfer of mass between sites
depend only on the departure site and not on the destination site.
It is natural to consider more general models in which
for example a hop rate takes the form $u(m,n)$ where $m$ is the occupancy of the departure site $i$ and $n$ is the occupancy of the destination site $i+1$. This type of model is variously referred
to as a misanthrope or  migration process.

\begin{figure}%
	\center
	\psfrag{m}{$m$} \psfrag{n}{$n$} \psfrag{umn}{$u(m,n)$}
	\includegraphics*[width=10.0cm]{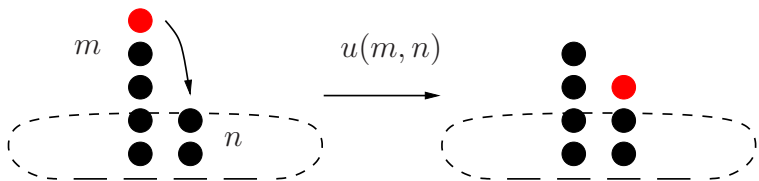}
	\caption{Definition of the Misanthrope process: a particle hops from site with $m$ to site with $n$ particles with rate $u(m,n)$.
	\label{model_def}}%
\end{figure}

We now define the model that we consider. As in the ZRP case, $M$ particles reside on sites of a 1D closed chain of length
$L$; each site $i$ carries $m_i$ particles, and the conservation of particles requires that $\sum_{i=1}^L
m_i=M$. The only difference with the ZRP is that a particle hops from site $i$ to site $i+1$ with rate
$u(m_i,m_{i+1})$ which depends on the occupancies of both the departure
and the arrival site, see Fig.~\ref{model_def}. 

This model has a factorized stationary state (\ref{fullfact}) when certain conditions on $u(m,n)$ 
are satisfied. Here, we simply quote the constraint on $u(m,n)$ without proof: 
\bq 
u(m,n) = u(m+1,n-1)
  \frac{u(1,m)u(n,0)}{u(m+1,0)u(1,n-1)} + u(m,0) -
  u(n,0). \label{miscond} 
\eq 
It can be shown that this relation actually reduces to two conditions: 
\ba u(n,m) = u(m+1,n-1)
  \frac{u(1,m)u(n,0)}{u(m+1,0)u(1,n-1)}, \label{c1} \\ 
  u(n,m) - u(m,n) = u(n,0)-u(m,0). \label{c2} 
\ea
These  conditions were first written down by Cocozza-Thivent \cite{cocozza-thivent_processus_1985}.
Under these conditions  the single site weights $f(m)$ obey a recursion
\bq 
f(n) =
  f(n-1) \frac{f(1)}{f(0)} \frac{u(1,n-1)}{u(n,0)} \label{fn}\;.  
\eq
Conditions (\ref{c1},\ref{c2}) uniquely define all rates $u(n,m)$ and
the  single-site weights $f(m)$ in terms of a set of basic hop rates
which we denote
\ba
u(m,0) = y_m, \label{ydef} \\
u(1,n) = x_n. \label{xdef}
\ea
Iterating Eq.~(\ref{fn}) and using the definitions of $y_m$, $x_n$ 
(\ref{ydef},\ref{xdef}) we obtain the following expression for the single-site weight  $f(n)$:
\bq
f(n) = f(0) \left(\frac{f(1)}{f(0)}\right)^n \prod_{i=1}^n 
\frac{x_{i-1}}{y_i}, \label{fngen}
\eq
in which, as noted previously, we may suppress the exponential factor $f(0) \left(\frac{f(1)}{f(0)}\right)^n$.

Equations (\ref{c1}) and (\ref{c2}) can be rewritten as two recursion 
relations which allow one to find $u(m,n)$ for $m>2,n>1$:
\ba
u(m+1,n-1) = u(n,m) \frac{y_{m+1}}{x_m} \frac{x_{n-1}}{y_n}, \label{c3} \\
u(n,m) = u(m,n) - y_m + y_n. \label{c4}
\ea
By iterating these equations one obtains unique expressions for all 
$u(m,n)$.
However, there is  an additional condition that $u(m,n)$ should be
non-negative for all $m,n$ for it to be a hopping rate. This imposes some constraints on $y_m,x_n$ which cannot be expressed in a closed form.
Therefore it remains an open problem to determine precisely which $x_n$ and $y_m$ lead
to a physical model with non negative  hopping rates. However, there exists a special case of $u(m,n)$ for which more progress has been made and we shall discuss it now.

\section{Factorized hop rate in the misanthrope process}
In  recent work \cite{waclaw_explosive_2012,evans_condensation_2014} we considered   the special case
\bq
u(m,n) = w(m) v(n) \label{ufac}
\eq
which corresponds to hopping rates whose dependence on departure and destination site factorizes.
One can check that for this form of the hop rate
equation (\ref{c1}) is automatically fulfilled. When $v(n)=\rm const$, 
equation (\ref{c2}) is also fulfilled and we recover the ZRP with 
$u(m,n)=w(m)$. When $v(n)\neq \rm const$, equation (\ref{c2}) leads to 
the relation between $w(n)$ and $v(n)$:
\bq
w(n) = C[v(n)-v(0)],
\eq
with some arbitrary, non-zero constant $C$. Thus
\bq
u(m,n) = C [v(m)-v(0)] v(n)  \label{ufact}
\eq
is the form of a factorized hop rate that yields a factorized steady state
(\ref{fullfact})
with the single-site weight given by
\begin{equation}
f(m) = \prod_{i=1}^m \frac{v(i-1)}{v(i)-v(0)}\;.
\end{equation}
Before we discuss the condition for condensation in this model, we shall briefly review two simple choices of $v(m)$ that lead to previously studied models.

\subsection{Partial Exclusion}
Our first simple example
of the factorized hopping rate (\ref{ufact})
is $C=-1$ and $v(n)=N-n$ with some integer $N>0$ we have
\bq
	u(m,n) = m(N-n)\;.
\label{pex}
\eq
If $N=1$ this reduces to the asymmetric simple exclusion process
where the occupancy of each site is limited to 1
and $u(1,1)=0$.
Similarly, the case of general integer $N >0$
 corresponds to `partial exclusion'
\cite{schutz_non-abelian_1994} where each site of a lattice contains at most $N$ particles.
In this context the rate (\ref{pex}) may be understood as each of $m$ particles  attempting hops forward to the next site
with rate one and the hopping attempt succeeding with probability $N-n$ where $n$ is the occupancy of the destination site.

\subsection{Inclusion Process}
If we  take $C=1$ and
\begin{equation}
v(n) = n+ d
\end{equation}
in (\ref{ufact}) where $d$ is a positive constant, we obtain
\begin{equation} 
u(m,n) = m(n+ d)\;.
\end{equation} 
This is the hop rate for the so-called Inclusion Process studied in
\cite{grosskinsky_condensation_2011}. In the limit $d\to 0$ this model exhibits a distinct form 
of condensation.

\section{Condensation in the misanthrope process with factorized hop rates}\label{sec:cond}
We are interested in a factorized form of $u(m,n)$ from Eq.~(\ref{ufact}) such that Eq.~(\ref{fn}) gives $f(n)\sim n^{-\gamma}$ with $\gamma>2$ which as we know leads to standard condensation. It turns out that 
the standard condensation can occur through two contrasting types of dynamics.
One mechanism  is through the hops rates decaying sufficiently slowly with $n$, $m$. This can be achieved through
\bq
v(m) \cong \beta \left(1-\frac{\alpha}{m}\right) \label{cc2}
\eq
which leads to 
\bq
u(m,n) \cong \beta (v(0)-\beta) - \frac{\alpha\beta (v(0)-\beta)}{n} + 
\frac{\alpha\beta^2}{m}, \label{umn_powlaw}
\eq
which decays with $m$ in a similar fashion to the ZRP case. The other mechanism is for the hop rates to {\em increase} with $m$, and $n$ as 
\bq
	u(m,n) \sim (mn)^\gamma
\eq
with $\gamma >2$, which is equivalent to 
\bq
v(m) \sim m^\gamma. \label{cc1}
\eq
We refer to the latter case as {\em explosive} condensation.
Explosive condensation exhibits strikingly different dynamical properties to the
ZRP-like condensation. In particular, the condensate emerges on a time scale which vanishes with system size $L$ as $\sim (\ln L)^{1-\gamma}$ for $\gamma>2$ \cite{waclaw_explosive_2012}. This is in contrast to the case (\ref{cc2}) or the ZRP case for which the time increases with $L$ as $\sim L^2$.

Interestingly, for the misanthrope process the {\em existence} of condensation depends not only on the asymptotic behaviour of $v(m)$ but also on $v(0)$.  This should be contrasted with the ZRP  for which it is  only the asymptotic decay of the hop rate  that determines condensation. 
To illustrate this point consider the case \cite{evans_condensation_2014}
\bq
	v(0)<1, \qquad v(m)=1+\frac{1}{m+1}. \label{ex2}
\eq
In this case one can find closed form expressions for the weights $f(n)$
and generating function $F(z)$
and one may determine the critical density
given by $z \to 1$:
\bq
\rho_c = 
\frac{4(1-v(0))}{3v(0) -2}.
\eq
We see  that $\rho_c \to 0$ as $v(0)\to 1$
and $\rho_c \to \infty$ as $v(0)\to 2/3$.
Conseequently   for $v(0)\leq 2/3$  there is no condensation, but for $2/3< v(0) <1$ there 
is standard condensation  and for $v(0)=1$ there is strong condensation, 
even though $v(m)$ is the same in all cases for $m>0$.



\section{Pair-factorized steady states}
So far we have discussed the processes in which the stationary
probability $P(m_1,\dots,m_L)$ factorizes over sites of a 1d closed
chain.  One can consider generalisations of this structure to, for
example, a pair-factorized state in which there is factorization over
pairs of adjacent sites in which the stationary probabilities take the
form (\ref{pairfact}) where $g(m,n)$ is the pairwise weight.
The factorized stationary state (\ref{fullfact}) is recovered when
$g(m,n)$  factorizes: $g(m,n) = a(m) b(n)$, in which case the single-site weight $f(m) = a(m) b(m)$. 

\begin{figure}
	\begin{center}
	\includegraphics[width=8.5cm]{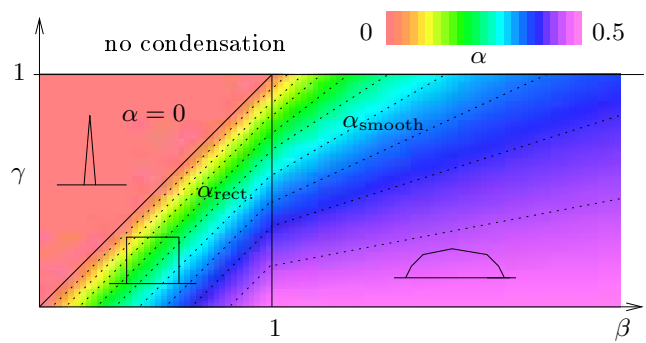}
	\end{center}
	\caption{\label{phase}Phase diagram for $K(x)\sim e^{-|x|^{\beta}}$ and $p(m)\sim e^{-m^\gamma}$. The spatial extension of the condensate $\sim L^\alpha$, where $\alpha_{\rm rect}=(\beta-\gamma)/(\beta-\gamma+1)$ for the rectangular and $\alpha_{\rm smooth}=(\beta-\gamma)/(2\beta-\gamma)$ for the smooth condensate. Dotted lines correspond to constant values of $\alpha=0.05,0.1,\dots,0.45$. Reproduced from \cite{waclaw_mass_2009}.}
\end{figure}

Such pair-factorized stationary states have been considered
in models \cite{evans_interaction-driven_2006,waclaw_tuning_2009} with  hopping rates  $u(m_{i-1},m_i,m_{i+1})$ which depend on the state of both (left and right) nearest neighbours:
\bq
	u(m_{i-1},m_i,m_{i+1}) = \frac{g(m_i-1,m_{i-1})}{g(m_i,m_{i-1})}\frac{g(m_i-1,m_{i+1})}{g(m_i,m_{i+1})}, \label{ratepffs}
\eq
where $g(m,n)$ is the same two-point weight that appears in the expression for the steady-state probability (\ref{pairfact}). It has been shown that a pair-factorized stationary state may modify the nature of condensation and allow a condensate spreading over a large, but non extensive number of sites. For example, if 
\bq
	g(m,n) = \exp\left[ -J |m-n| + (U/2) (\delta_{m,0} + \delta_{n,0})\right],
\eq
the condensate's shape is a distorted parabola extending to $\sim \sqrt{L}$ sites \cite{evans_interaction-driven_2006}.
References \cite{waclaw_tuning_2009,waclaw_mass_2009} have considered a more general case
\bq
	g(m,n) = K(|m-n|) \sqrt{p(m)p(n)}, \label{gkpp}
\eq
where $K(m)$ and $p(m)$ can be arbitrary, sufficiently-fast decaying functions. It turns out that when 
\bq
	K(x)\sim e^{-a|x|^\beta}, \qquad p(m)\sim e^{-b m^\gamma}, \label{expab}
\eq
for $a,b,\beta,\gamma>0$, condensation occurs above a certain critical density of particles if $\gamma<1$. The shape of the condensate changes from a single-site one, through a rectangular condensate, to a parabolic condensate as $\beta$ increases from zero to one, and from one to infinity, see the phase diagram in Fig. \ref{phase}. The scaling of the width of the condensate with $L$ depends on the parameters $\beta, \gamma$ in a non-trivial way. These results have been recently confirmed numerically \cite{ehrenpreis_numerical_2014}, with some small discrepancies attributed to finite-size effects.

\section{Pair-factorized states for discrete time dynamics}
The pair-factorized steady state discussed in the previous section assumes a three-site hop rate (\ref{ratepffs}). A very interesting question to ask is whether a two-point $u(m,n)$ such as the one in the misanthrope process can also lead to a pair-factorized steady state.

One might  have hoped that when the hop rate $u(n,m)$ does not satisfy
the conditions for a factorized stationary state, there would still be
some choices of rates $u(m,n)$ which yield pair-factorized
states. However it was shown in \cite{evans_condensation_2014} that this is not the case. Therefore  the misanthrope process either  has a factorized stationary state or the stationary state has some unknown structure.

Here we consider a generalisation of the misanthrope process similar
to the generalisation of the zero range process reviewed in
Section~\ref{Sec:ZRPgen}.  That is, we consider a Misanthrope-like
model with stochastic discrete-time parallel dynamics where all
particles attempt hops at discrete time steps.  Our aim is determine
how the structure of the stationary state is modified.  It turns out that we
need to consider pair factorization for the steady state probability.

The dynamics we consider generalise the misanthrope process in two ways. First
masses move at discrete timesteps therefore there can be
simultaneous transport of mass at different locations.  Second, more
than one unit of mass can be transferred from site $i$ to $i+1$.

More precisely,  in each time step, some number
$\mu_i$ of particles depart from site $i$ and move to site $i+1$ with
probability $\phi(\mu_i|m_i,m_{i+1})$.  Clearly, we must have
$\phi(\mu|m,n)=0$ for $\mu>m$ and for conservation of probability we
require $\sum_{\mu=0}^m \phi(\mu|m,n)=1$.  Following \cite{evans_factorized_2004} we will refer to 
$\phi(\mu| m,n)$ as the chipping kernel.

One would like to know whether there is a  necessary and sufficient condition on  $\phi(\mu|m,n)$ for the steady state $P(m_1,\dots,m_L)$ to factorize over pairs of neighboring sites. This means that the
following  equation must be fulfilled:
\ba
	\prod_{i=1}^L g(m_i,m_{i+1}) &=& \sum_{\mu_1,\dots,\mu_L} \prod_{i=1}^L \phi(\mu_i|m_i+\mu_i-\mu_{i-1},m_{i+1}+\mu_{i+1}-\mu_i) \nonumber \\
	& & \times g(m_i+\mu_i-\mu_{i-1},m_{i+1}+\mu_{i+1}-\mu_i)\;. \label{ss}
\ea
The left hand side of this equation gives the weight (unnormalized probability) of some configuration
of mass in the system  given by $\{ m_i\}$. The right hand side  gives the sum over the weights of possible configurations at the previous time step
multiplied by the transition probabilities to the  configuration
$\{ m_i\}$. The sum over preceding configurations is expressed as a sum over the masses, $\mu_i$,  transferred from $i$ to $i+1$.
Thus at the previous time step site $i$ had mass $m_i+\mu_i -\mu_{i-1}$.
For a stationary state the two sides of (\ref{ss}) must be equal.

We now claim that a chipping kernel of the form
\bq
	\phi(\mu|m,n) = \frac{u(\mu) a(m-\mu) b(n+\mu)}{g(m,n)}, \label{nufact}
\eq
where
\begin{equation}
g(m,n)=\sum_{\mu=0}^m u(\mu)a(m-\mu)b(n+\mu),
\label{gdef}
\end{equation}
is sufficient for pair factorization provided that some additional constraint is imposed on the functions $u(\mu),a(m),b(n)$. 
Expression (\ref{nufact}) implies a factorization of the chipping kernel
into a factor $u(\mu)$ which depends on the mass transferred, a factor $a(m-\mu)$ which depends on the mass which remains at the departure site
and a factor $b(n+\mu)$ which depends on the resulting mass  at the 
destination site.

To demonstrate that the form  (\ref{nufact}) indeed leads to pair-factorized
stationary state, let us insert (\ref{nufact}) into the stationarity condition Eq.~(\ref{ss}). Rearranging indices we obtain
\ba
	\prod_i g(m_i,m_{i+1}) &=& \sum_{\{\mu_i\}} \prod_i \left(  u(\mu_i) a(m_{i}-\mu_{i-1}) b(m_{i+1}+\mu_{i+1}) \right) \nonumber\\
&=& \prod_i \left( \sum_{\mu_i} u(\mu_i) a(m_{i+1}-\mu_i) b(m_{i}+\mu_i) \right) \nonumber\\
	&=& \prod_i g(m_{i+1},m_i)\;.
\ea
In other words, for pair factorization under the form (\ref{nufact}) we require
\bq
	\prod_i g(m_i,m_{i+1})/g(m_{i+1},m_i) = 1\;.
\eq
The most general solution to this  equation
assumes the form (for a proof see \cite{evans_condensation_2014} Appendix A) 
\bq
	g(m,n) = g(n,m) \frac{h(n)}{h(m)}, \label{gcond}
\eq
where $h(m)$ is some function which can be determined by inserting $n=0$ into the above equation:
\bq
	h(m)=h(0) \frac{g(0,m)}{g(m,0)}\;. \label{hcond}
\eq
Finally, from (\ref{gcond}) and (\ref{hcond}) we obtain 
\bq
	g(m,n)g(0,m)g(n,0) = g(n,m) g(m,0)g(0,n)\;. \label{gcond2}
\eq
Equation (\ref{gcond2}), in tandem with the definition (\ref{gdef}), is the central result of this section and gives a sufficient condition for the stationary state of the generalised misanthrope process to take a pair-factorized form. 
It implies conditions on
 $u(\mu),a(m),b(n)$ which we explore in the next section. It remains an open problem whether Eq. (\ref{gcond2})  is also a necessary condition.

\subsection{Single particle hopping}
To simplify the discussion we 
we now focus on the misanthrope-like case when at most one particle can hop from a given site in each timestep.  For this purpose we take  $u(\mu)=\delta_{\mu,0}+\dt \delta_{\mu,1}$, and (\ref{gdef}) yields
\bq
	g(m,n) = a(m)b(n) + \dt a(m-1)b(n+1). \label{gmn}
\eq
At this stage $\Delta t$ is a parameter but
as we shall verify later the limit $\Delta t \to 0$  reduces to the usual continuous time  misanthrope process.
We will derive now the relation between $a(m)$ and $b(m)$ for arbitrary $\dt>0$. 
Inserting (\ref{gmn}) in condition (\ref{gcond2}) leads, after some algebra, to
\bq
	\frac{\bt(0)-\bt(m) }{\at(m)} - \dt\bt(0) \bt(m) = 
	\frac{\bt(0)-\bt(n) }{\at(n)} - \dt\bt(0) \bt(n), \label{eqtilde}
\eq
where we have defined 
\ba
	\at(m)=a(m-1)/a(m), \\
	\bt(m)=b(m+1)/b(m).
\ea
The left and right side of  equation (\ref{eqtilde}) are functions of $m$ and $n$, respectively. In order for (\ref{eqtilde}) to be valid for any $m$, $n$, both sides of the above equation must be equal to a constant $k$. This gives
\bq
	\bt(m) = \frac{\bt(0)-\frac{1}{k}\at(m)}{1+\dt\bt(0)\at(m)} \label{tbm}
\eq
or, equivalently,
\bq
	\at(m) = k \frac{\bt(0)-\bt(m)}{1+k \dt \bt(0)\bt(m)}\;. \label{tam}
\eq
Therefore, $\at(m)$ is determined by $\bt(m)$ and vice versa, up to two parameters $k$ and $\dt$. The chipping probability $\phi(1|m,n)$ can  then be expressed as
\bq
	\phi(1|m,n) = k\dt \bt(n) \frac{\bt(0)-\bt(m)}{1+k\dt (\bt(m)\bt(0) + \bt(n)\bt(0) - \bt(m)\bt(n))},
\eq
and because the hopping probability defines the model, all static and dynamical properties are fully specified by giving $k,\dt$ and one of the two functions $\at(m)$ or $\bt(m)$. The pairwise weight function $g(m,n)$ is given by Eq.~(\ref{gmn}), with $a(m),b(n)$ calculated recursively:
\bq
	a(m) = \prod_{i=1}^m \frac{1}{\at(i)}, \qquad b(n) = \prod_{i=1}^n \bt(i-1) ,
\eq
where we assumed for convenience that $a(0)=b(0)=1$. This assumption is made without loss of generality
as it only rescales $g(m,n)$ by a constant factor.

In the limit $\dt\to 0$, we obtain  factorization of the steady state, since (\ref{gmn}) tends to $g(m,n)=a(m)b(n)$ corresponding to a single-site weight $f(m) = a(m) b(m)$.
The chipping probability becomes
\bq
	\phi(1|m,n) = \Delta t \, \at(m) \bt(n) 
+ O(\Delta t^2),
\eq
and in the continuous time limit $\Delta t \to 0$ this reduces to a  hopping rate
\bq
	u(m,n) = \at(m) \bt(n) ,
\eq
whereas condition (\ref{eqtilde}) reduces to
\bq
	 \frac{\bt(0)-\bt(m)}{\at(m)}= \frac{\bt(0)-\bt(n)}{\at(n)} ,
\eq
which is equivalent to the condition (\ref{c2}) for the misanthrope process.

\subsection{Condition for condensation}

In this section we shall outline how the above discrete-time misanthrope process with factorized hopping probability exhibits condensation above some critical density of particles. We shall learn, perhaps not surprisingly, that the conditions  on the hopping probability $\phi(1|m,n)$ are somewhat  similar to those for the hopping rate $u(m,n)$ in the continuous time case.

We begin by defining the grand-canonical partition function
\bq
	G(z) = \sum_{\{m_i\}} z^{\sum_i m_i} W(\{m_i\}),
\eq
where the steady-state weights reads
\ba
	W(\{m_i\}) = \prod_{i=1}^L g(m_i,m_{i+1}) \nonumber \\
	=  \prod_{i=1}^L\left[ a(m_i)b(m_{i+1}) + \dt a(m_i-1)b(m_{i+1}+1) \right].
\label{Wpfac}
\ea
We now observe that the expression in square brackets in Eq. (\ref{Wpfac}) can be viewed as a product of two vectors:
\bq
	\left( a(m_i) + \dt a(m_i-1) \right) 
	\left(\begin{array}{c} b(m_{i+1}) \\ b(m_{i+1}+1) \end{array} \right),
\eq
and hence the steady-state weight can be rewritten as
\ba
	W(\{m_i\}) &=& \prod_{i=1}^L  \left( a(m_i) + \dt a(m_i-1) \right)\left(\begin{array}{c} b(m_{i+1}) \\ b(m_{i+1}+1) \end{array} \right) \nonumber \\
	&=& \Tr \left[ \prod_{i=1}^L \left( a(m_i) + \dt a(m_i-1) \right)\left(\begin{array}{c} b(m_{i+1}) \\ b(m_{i+1}+1) \end{array} \right) \right] \nonumber \\
	&=& \Tr \left[ \prod_{i=1}^L \left(\begin{array}{c} b(m_{i}) \\ b(m_{i}+1) \end{array} \right)\left( a(m_i) + \dt a(m_i-1) \right) \right] \nonumber \\
	&=& \Tr \left[ \prod_{i=1}^L \left( \begin{array}{cc} 
		a(m_i)b(m_i) & \dt a(m_i-1)b(m_i) \\
		a(m_i)b(m_i+1) & \dt a(m_i-1)b(m_i+1) 
	\end{array}\right) \right],
\ea
where in the penultimate step we have cyclically permuted the vectors under the trace and in the last step evaluated the resulting dyadic product. The grand-canonical partition function becomes
\bq
	G(z) = \Tr \left[ A(z)^L\right],
\eq
where $A(z)$ is a $2\times 2$ matrix:
\bq
	A(z) = \left( \begin{array}{cc} 
		\sum_m a(m)b(m)z^m & \dt \sum_m a(m-1)b(m)z^m \\
		\sum_m a(m)b(m+1)z^m & \dt \sum_m a(m-1)b(m+1)z^m
	\end{array}\right). \label{Az}
\eq
The above matrix has two eigenvalues
\bq
	\lambda_{\pm}(z) = \frac{1}{2} \left(A_{11}(z)+A_{22}(z) \pm \sqrt{(A_{11}(z)-A_{22}(z))^2 + 4 A_{12}(z) A_{21}(z)} \right), \label{lpm}
\eq
where $A_{ij}$ is an element of (\ref{Az}). 
Since $\lambda_+$ is always greater than $\lambda_-$, provided that $z>0$, 
$\ln G(z) \cong L \ln \lambda_+$ for large system size $L$, and we obtain that, in the thermodynamic limit, the density-fugacity relation is given by
\bq
	\rho(z) = z\frac{\lambda^{'}_+(z)}{\lambda_+(z)}\;.
\eq
Since $\rho(z)$ is an increasing function of $z$, condensation will
occur if $\rho(z)$ tends to a finite value  as $z$ approaches
its maximum allowed value. This maximum allowed values is $z_c$ -- the radius of convergence of $\lambda_+(z)$. From Eqs.~(\ref{lpm}) and (\ref{Az}) we see that the radius of convergence
of $\lambda_+(z)$ is the radius of convergence of any of the four entries of $A(z)$. If we focus, say, on $A_{11}(z)$, we see that condensation criteria reduce to the convergence properties
of the sum $\sum_m a(m)b(m)z^m$.
Hence,  condensation is possible if
\bq
	\tilde f(m)=a(m)b(m)
\eq
behaves in one of two ways described earlier
i.e.  here $\tilde f(m)$ plays the same role as the single-site weight $f(m)$ 
in section~\ref{intro} cases {\bf I}, {\bf II}.


\section{Discussion}
In this paper we have given a short review of condensation in
factorized and pair-factorized states. In particular we have discussed
the misanthrope process where the hop rates $u(m_i,m_{i+1})$ depend on
the occupancy of both the departure site $i$ and destination site
$i+1$.  This process provides a new route to the standard condensation
scenario for the case of increasing hop rate $u(m,n)\sim m^\gamma
n^\gamma$ for $\gamma >2$.

We have studied a generalisation of the misanthrope process to a
parallel discrete time updating scheme in which simultaneous transfer
of mass at different locations can occur. We have shown that
conditions exist for the stationary state to take a pair-factorized
form.  These conditions generalize the conditions (\ref{c1},\ref{c2})
for factorization in the continuous time case.  Thus the continuous
time factorized stationary state is modified into a pair-factorized
state when discrete time dynamics is considered. This contrasts with
the zero-range process in which a factorized form is maintained under
discrete time updating.

A major open question which remains is the structure of the
misanthrope process stationary state for general hopping rates. Also
it would be of interest to generalise the conditions for factorization
and pair factorization in the misanthrope process to more general
geometries than the one dimensional periodic chain.

\section*{Acknowledgments}
We would like to thank  David Mukamel for helpful discussion. M.R.E. thanks GGI, Firenze, for hospitality. B.W. thanks the Leverhulme Trust and the Royal Society of Edinburgh for support. This work was funded in part by the EPSRC under grant number EPSRC J007404. 

\section*{References}

\bibliographystyle{iopart-num}
\bibliography{references}

\end{document}